\documentstyle[12pt,aps,prb]{revtex}
\begin{document}
\author{V.Ya.Demikhovskii\footnote{Corresponding author.
Mail: Russia, 603006, Nizhny Novgorod, M.Gorky St., 156-3;
E-mail: demi@phys.unn.runnet.ru; FAX: 007 8 312 658592},
A.A.Perov, D.V.Khomitsky}
\address{Nizhny Novgorod State University\\
RUSSIA, 603600 Nizhny Novgorod, Gagarin Ave., 23}
\title{Formation of New Fermi Surfaces in 3D Crystals at Ultra High
Magnetic Field with Different Orientations}
\maketitle

\vspace{1cm}
\begin{abstract}
In the tight-binding approximation the Harper like equation describing an
electron in 3D crystal subject to a uniform magnetic field is obtained. It
is supposed that the vector ${\bf H}$ can be oriented along several directions
in the lattice. The Fermi surfaces relevant to a magnetic flux $p/q=1/2$ in a
simple cubic lattice are built. The quantization rules in magnetic fields
slightly distinguished from $p/q=1/2$ are investigated.
\end{abstract}

\vspace{1cm}
{\bf PACS} number(s):~~~~71.18.+y, 71.20.-b


\vspace{1cm}
The effects of a high magnetic field on Bloch electrons are fascinating
problem with very rich physics. This is not surprising since there are two
fundamental periods in the problem -- period of the potential and the period
of the phase of the wave function -- and the interplay of these
two periods gives very interesting spectrum (Hofstadter butterfly) and
eigenstate structure. Starting from classical works of Azbel [1],
Hofstadter [2], Wannier [3] (see also [4-7]) this problem attracts the
increasing attention. During the last decade the one electron
quantum states in lateral superlattices in the presence of perpendicular
magnetic field have been studied in a series of theoretical [8-10] and
experimental [11,12] works. But unfortunately, all works mentioned above
deals only with 2D crystals subject to a perpendicular magnetic field.
Nowadays when in VNIIEF (Sarov) the magnetic field up to 28 MGs is reached
[13] the idea of observation of details forecasted by the theory in the
crystals with the lattice spacing of $a=0.6~nm$ and more does not seem to
be unaccessible.  

In this paper we shall consider electronic states in 3D crystals subject to
ultrahigh magnetic field. We consider that the vector H is parallel to
arbitrary translation vector of a lattice. In a tight-binding approximation
the wave function satisfying the generalized Bloch-Peierls conditions is
obtained and Harper like equation for different orientations of a magnetic
field is derived. As an example, the Fermi surfaces for magnetic subbands relevant
to rational value of a magnetic flux $p/q=1/2$ are built. The quantization rules for
magnetic flux near this rational value of flux quanta are studied.

First of all in the tight-binding approximation let us derive Harper's
equation for the simple cubic lattice subject to uniform magnetic field.
To be specific let the basis vectors on the $x,y$ plane be
${\bf a}_1$ and ${\bf a}_2$, and the magnetic field is applied
along lattice basis vector ${\bf a}_3\parallel 0z$ ($|{\bf a}_1|=
|{\bf a}_2|=|{\bf a}_3|=a$). Let vector potential is be chosen in Landau
gauge ${\bf A}=(0,H\,x,0)$. Then electron wave function must satisfy the
generalized Bloch conditions (Peierls conditions) [14]
$$
\psi_{{\bf k}}({\bf r})=\psi_{{\bf k}}(x+qa,y+a,z+a)\exp(-ik_xqa)
\exp(-ik_ya)\exp(-ik_za)\exp(-2\pi ipy/a),\eqno(1)
$$
due to the fact that the vector potential is not a periodic function of the
coordinates. Here $p$ and $q$ are mutually prime integers which define the
number of magnetic flux quanta per two-dimensional square elementary cell
$$
\frac{p}{q}=\frac{\Phi}{\Phi_0}=\frac{|e|H|[{\bf a}_1\times{\bf a}_2]|}
{2\pi\hbar c},\eqno(2)
$$
where $\Phi_0=hc/|e|$ is the magnetic flux quanta.
In a presence of a magnetic field the electron quantum states must be
classified in accordance with irreducible representations of magnetic
translation group
$$
{\bf a}_{mag}=nq{\bf a}_1+m{\bf a}_2+l{\bf a}_3, \quad (n,m,l{\rm ~are~
integer~numbers}).
$$
The quasimomentum ${\bf k}$ takes values in the magnetic Brillouin zone
$$
-\frac{\pi}{qa}\le k_x\le \frac{\pi}{qa},\quad
-\frac{\pi}{a}\le k_y\le \frac{\pi}{a},\quad
-\frac{\pi}{a}\le k_z\le \frac{\pi}{a}.\eqno(3)
$$

In the tight-binding approximation the electron wave function which satisfies
the conditions (1) can be written in the form [15]
$$
\psi_{{\bf k}}({\bf r})=\sum_{n,m,l}g_n({\bf k})\exp(i{\bf ka_n})
\exp\Bigg(-2\pi i\frac{p}{q}\frac{(y-ma)}{a}n\Bigg)\psi_0(x-na,y-ma,z-la)
\eqno(4)
$$
where $\psi_0({\bf r}-{\bf a_n})$ is the atomic function. The path-dependent
geometric phase $2\pi\frac{p}{q}\frac{(y-ma)}{a}n$ of the wave function is
known to play a fundamental role in the problem.

By substituting of (4) into Schr\"odinger equation and evaluating the
transfer integrals between neighbour sights we shall obtain a system of
difference equations for coefficients $g_n$. The magnetic field has a
non-trivial influence on the transfer integrals between neighbour lattice
sights in the $x$ direction
$$
A=\int\exp\Bigg(\pm2\pi i\frac{p}{q}\frac{y-ma}{a}\Bigg)\psi_0({\bf r}-
{\bf a_n^{\prime}})
(V({\bf r})-U({\bf r-a_n}))\psi_0({\bf r}-{\bf a_n})\, d\tau,\eqno(5)
$$
where ${\bf a_n^{\prime}}=\Big((n\pm 1)a,ma,la\Big)$,
$V({\bf r})$ is the crystal scalar potential with periodicity $a$ in three
dimensions and $U$ is the atomic potential. Calculating the
transfer integrals in the mean-value approximation we substitute $y=ma$ and
obtain $A=E_0$ where $E_0$ is the transfer integral (5) in the absence of a
magnetic field. In the $y$ direction the transfer integral is
$$
B=\exp\Bigg(\pm2\pi i\frac{p}{q}n\Bigg)\int\psi_0({\bf r}-{\bf a_n^{\prime}})
(V({\bf r})-U({\bf r-a_n}))\psi_0({\bf r}-{\bf a_n})\, d\tau,\eqno(6)
$$
where ${\bf a_n^{\prime}}=\Big(na,(m\pm 1)a,la\Big)$ and in the $z$ direction
the transfer integral is equal to $E_0$.
As a result we obtain the well-known Harper's equation [4]
$$
\exp(ik_xa)g_{n+1}+\exp(-ik_xa)g_{n-1}+2g_n\cos\Big(2\pi n\frac{p}{q}+k_ya
\Big)=\varepsilon(k_x,k_y,k_z)g_n,\eqno(7)
$$
where 
$$
\varepsilon(k_x,k_y,k_z)=\varepsilon_{\perp}(k_x,k_y)-2\cos k_za\eqno(7a)
$$
is the dimensionless energy measured in terms of $E_0$.
The spectrum $\varepsilon_{\perp}(k_x,k_y)$ in Eq.(7) consists of $q$ bands
and depend on one single parameter $p/q$ counting the number of magnetic
flux quanta per unit cell (Fig.1). If flux is irrational the spectrum is a
singular-continuum -- uncountable but measure zero set of points (Cantor set).
Originally the Eq.(7) was derived by Harper [4] using Peierls
substitution [14].

For some values of $p/q=1/2,\,1/3,\,1/4$ it is possible to solve the Eq.(7)
analytically. As a result one can obtain the following expressions for
dispersion laws for $q$ magnetic subbands: for $p/q=1/2$ (two subbands)
$$
\displaylines{\varepsilon^{(1,2)}_{\perp}(k_x,k_y)=
\pm 2\sqrt{\cos^2k_xa+\cos^2k_ya},\hfill\llap{(8)}\cr}
$$
for $p/q=1/3$ (three subbands)
$$
\displaylines{\varepsilon^{(1)}_{\perp}(k_x,k_y)=-2\sqrt{2}\cos\Big(\frac{2\pi}{3}
+(\arctan\beta)/3\Big),\hfill\cr
\varepsilon^{(2)}_{\perp}(k_x,k_y)=2\sqrt{2}\cos\Big(\frac{\pi}{3}
+(\arctan\beta)/3\Big),\hfill\cr
\varepsilon^{(3)}_{\perp}(k_x,k_y)=-2\sqrt{2}\cos
\Big((\arctan\beta)/3\Big),\hfill\llap{(9)}\cr}
$$
where $\beta=\sqrt{8-\alpha^2}/\alpha,\,\alpha=\cos 3k_xa+\cos 3k_ya$; and
for $p/q=1/4$ (four subbands)
$$
\displaylines{\varepsilon^{(1,2,3,4)}_{\perp}(k_x,k_y)=
\pm\sqrt{4\pm 2\sqrt{4-\sin^22k_xa-\sin^22k_ya}}.\hfill\cr}
$$

Fig.2 shows the Fermi surfaces in the lowest subband for $p/q=1/2$ obtained
using (7a) and (8) in the first Brillouin zone $-\pi/qa\le k_x\le
\pi/qa,\;-\pi/qa\le k_y\le\pi/qa,\;-\pi/a\le k_z\le\pi/a$. Three surfaces for
different representative energies are plotted. For an ellipsoid-type surface
such as (1) ($|{\bf k}|\to 0$) the simple analytical expression for an
energy spectrum $\varepsilon(\bf k)$ can be easily obtained with the help of
(7a) and (8). The relevant effective masses are determined as
$$
m^{*}_{x,y}=\frac{\hbar^2}{\sqrt{2}E_0a^2},\; m^{*}_z=\frac{\hbar^2}
{2E_0a^2}.\eqno(11)
$$
The asymptotic spectrum near the point $k_{x,y}\to\pi/2a$ ((2)-type surface)
is $\varepsilon(k_x,k_y,k_z)\sim\sqrt{k_x^2+k_y^2}-2\cos k_z$.

Now we shall discuss the magnetic quantization rules at values of a flux
slightly distinguished from $p/q=1/2$:
$$
\frac{p}{q}=\frac{1}{2}+\frac{1}{q^{\prime}},\quad q^{\prime}\gg 1.\eqno(12)
$$
This corresponds to the magnetic field $H=H(1/2)+\Delta H(1/q^{\prime})$.
As one can see from Fig.1 near value of $p/q=1/2$ the spectrum
represents a system of narrow subbands (practically discrete levels). Here it
is possible to see an equidistant spectrum, points of level accumulation and
ranges where $\varepsilon_{\perp}\sim\sqrt{N}$, ($N$ is the level number).

The spectrum $\varepsilon_{\perp}(k_x,k_y)$ of Harper's equation near one
half of flux quanta can be studied analytically due to the self-similarity
of its structure. In the case $p/q=1/2$ two subbands and two subsystems in
the eigenvector distribution can be observed. It may be derived analytically
from Harper's equation that every eigenvector function from a subsystem has
an argument step $\Delta n=2$.

The magnetic flux (12) gives us the diagonal term in Harper's equation of
the form $2\cos(\pi n+2\pi n/q^{\prime})g_n=2\cos\pi n\cos(2\pi n/q^{\prime})
g_n$. Now $q^{\prime}$ is a new period of the equation. Since
$\cos\pi n=(-1)^n$ our system of equations (7) consists of two subsystems where only
the sign varies. It is easy to construct an equation for each subsystem. After
simple algebra we obtain $(k_x=k_y=0)$:
$$
g_{n+2}+g_{n-2}+2\cos\frac{4\pi}{q^{\prime}}n\,g_n=
(\varepsilon^2_{\perp}-4)g_n.\eqno(13)
$$
Now $\varepsilon^2_{\perp}-4=\varepsilon_{\perp}^{\prime}$ is the energy
in Harper's equation (7) and $\varepsilon_{\perp}=
\pm\sqrt{\varepsilon_{\perp}^{\prime}+4}$ describes the structure of the
spectrum near one half of flux quanta and $\cos k_za=0$. Two signs at the
square root correspond to two groups of energy subbands. The energy
$\varepsilon_{\perp}$ describes all parts of the spectrum near $p/q=1/2$
(Fig.1). We can obtain from (13) the "square root" dependence at
$\varepsilon_{\perp}=0$, the clustering points $\varepsilon_{\perp}=\pm 2$
and Landau levels near $\varepsilon_{\perp}=\pm 2\sqrt{2}$. These peculiarities
of a spectrum can be obtained with the help of Onsager-Lifshitz quasiclassical
quantization rules on the Fermi surfaces (Fig.2a,b). It is easy to see that
magnetic quasiclassical quantization on the (1)-type surface results to the
eqidistant energy spectrum at the bottom of the lowest magnetic subband.
Accordingly, the magnetic Onsager-Lifshitz quantization near $k_z\to 0$ on
(2)-type surface yields the "square root" spectrum. At last, it is clear that
in magnetic field ${\bf H\parallel 0z}$ there are self-crossing orbits on
(3)-type surface. It gives the accumulation point in the spectrum
$\varepsilon_{\perp}$ near $p/q=1/2$ (see Fig.1). It can be stressed that
level spacing obtained from (13) has the same value that was derived using
quasiclassical quantization.

In accordance with Harper's equation (7) and (7a) the partial overlapping
between $q$ energy subbands is observed. When $p/q=1/2$ the overlapping takes
place in the energy interval $(-2,\,2)$. It is clear that when the electron
concentrations leading to $\varepsilon_{F}\in (-2,\,2)$ in
the magnetic field $H=H(1/2)+\Delta H(1/q^{\prime})$ the de Haas - van Alfen
oscillations with different periods corresponding to the different energy
level series (Fig.1) can be observed.

Further it is necessary to make the following relevant note.
The wave function (4) describes non-homogeneous probability
distribution on the lattice sights. This is due to the amplitudes $g_n({\bf k})$
which give us non-uniform shapes. However, in real crystals due to Coloumb 
interaction the static electron density distribution must maintain 
its symmetry even in the presence of magnetic field. 

Fortunately, the spectrum of Harper's equation is $q$-fold degenerate since
$\varepsilon(k_x,k_y+2\pi j/qa,k_z)=\varepsilon(k_x,k_y,k_z),\;
j=0,1,2,\dots,q-1$ and a wave function with homogenious electron density
distribution can be written as a combination of (4). Let us introduce
$$
\psi^{\prime}_{{\bf k}}({\bf r})=\sum_{j=0}^{q-1}C_j({\bf k})\hat K_j
\hat T_{ja}\psi_{{\bf k}}({\bf r}),\eqno(14)
$$
where $\psi_{{\bf k}}({\bf r})$ is from (4), $C_j({\bf k})=\exp(-ik_xja)$,
$\hat T_{ja}\psi=\psi(x+ja,y,z)\exp(ief/\hbar c)$ ($f=-Hyja$),
the operator $\hat K_j$ transforms $\psi_{{\bf k}}$ with
${\bf k}=(k_x,k_y,k_z)$ into $\psi_{{\bf k}^{\prime}}$ with
${\bf k}^{\prime}=(k_x,k_y-2\pi pj/qa,k_z)$. After this procedure we can
write (14) in the form
$$
\psi_{{\bf k}}^{\prime}({\bf r})=D({\bf k})\sum_{{\bf n}}\exp\Bigg(-2\pi i
\frac{p}{q}\frac{(y-ma)}{a}n\Bigg)\exp(i{\bf ka_n})\psi_0({\bf r-a_n}),
\eqno(15)
$$
where $D({\bf k})=\sum_{j=0}^{q-1}g_{n+j}({\bf k})$ does not depend on $n$
due to the periodicity of the system. Now (15) describes homogeneous
electron density distribution on the lattice sights and corresponds to
the same energy as $\psi_{{\bf k}}$.

Let us consider now the case of a simple cubic lattice subjected to a uniform
magnetic field oriented along the diagonal of a square in the plane $(x,y)$.
The magnetic field with amplitude $H$ has now the following Cartesian
coordinates:
$$
{\bf H}=\frac{H}{\sqrt{2}}(1,1,0).\eqno(16)
$$
It is convenient to choose a new coordinate system making a rotation about
the old $z$ axis by $\pi/4$ and defining the magnetic field orientation as
$x_3$. This transformation is written as following:
$$
\displaylines{\hfill x=\frac{x_3-x_1}{\sqrt{2}},\hfill\cr
\hfill y=\frac{x_3+x_1}{\sqrt{2}},\hfill\cr
\hfill z=x_2.\hfill\cr}
$$

We choose a new elementary cell: it is a rectangular parallelepiped based
on the new coordinate system vectors: a square with the side $\sqrt{2}a$ at
the base in the plane $(x_1,x_3)$ and the height $a$. This cell is non-primitive:
it consists of two atoms (additional sights are located at the centers of top
and bottom faces). In this coordinate system the magnetic field is oriented
along one of the elementary cell basic vectors (namely, ${\bf a}_3$).
According to the general principles in this case the classification of
electron states is possible [16]. Now the first Brillouin zone has an area
two times less with respect to the initial value.
The magnetic field in this coordinate system is written:
$$
{\bf H}=H(0,0,1)\eqno(17)
$$  
and it is convenient to choose a vector potential using Landau gauge:
$$
{\bf A}=(-H\, x_2,0,0).\eqno(18)
$$

The wavefunction in the tight-binding approximation can be written
as following:
$$
\displaylines{\hfill\psi_{{\bf k}}({\bf r})=\sum_{{\bf n}}g_n({\bf k})
\Bigg\{\exp\Big(2\pi i\frac{p}{q}\frac{x_1-m\sqrt{2}a}
{a/\sqrt{2}}n\Big)\exp(i{\bf ka_n})\psi_0({\bf r-a_n})+\hfill\cr
\hfill +\exp\Big(2\pi i\frac{p}{q}\frac{x_1-(m+1/2)\sqrt{2}a}
{a/\sqrt{2}}n\Big)\exp(i{\bf k(a_n+d)})\psi_0({\bf r-(a_n+d)})\Bigg\}.
\hfill\llap{(19)}\cr}
$$
Here ${\bf ka_n}=k_1\sqrt{2}am+k_2an+k_3\sqrt{2}al$, $\psi_0$ is an isolated
atomic function in the presence of the magnetic field, the amplitudes
$g_n({\bf k})$ are the subject of further research and $g_{n+q}=g_n$.
The indexes $(m,n,l)$ are taking all integer values independently,
$n$ numerates sights along $x_2$ axis, perpendicular to the magnetic field.
The sum is taken over all lattice sights and two addendums in (22)
correspond to the two existing crystalline subsystems. Each of them is a
simple tetragonal lattice with the elementary cell being a rectangular
parallelepiped with square of the side $\sqrt{2}a$ in its base and the
height $a$. These subsystems are displaced on the vector ${\bf d}$ with
new coordinates $(a/\sqrt{2},0,a/\sqrt{2})$. The magnetic translation group
is constructed by the $q$ - time multiplication of the translation period
along $x_2$ axis: $x_2\to x_2+qa$. The other translations are not changed.
Their periods in the new coordinate system are $\sqrt{2}a$ which is the
diagonal of a square with the side $a$. The classification of wavefunctions
is possible when the number of flux quanta $\Phi/\Phi_0$ through the area
$a^2/\sqrt{2}$ (being the minimal inter-sight area in the transversal to
magnetic field projection) is a rational number: $\Phi/\Phi_0=p/q$ where $p$
and $q$ are mutually prime integers.

The substitution of (19) into Schr\"odinger equation is performed and the
transfer integrals are calculated. We will have the equation for
the energy $\varepsilon$ and the amplitudes $g_n({\bf k})$:
$$
e^{ik_2a}g_{n+1}+e^{-ik_2a}g_{n-1}+4\cos\Big(\frac{k_3a}{\sqrt{2}}\Big)
\cos\Big(2\pi\frac{p}{q}n+\frac{k_1a}{\sqrt{2}}\Big)g_n=\varepsilon g_n.
\eqno(20)
$$
It should be mentioned that this equation may be obtained using Peierls
substitution ${\bf k}\to \hat k-\frac{e{\bf A}}{\hbar c}$ with the initial
spectrum at null magnetic field. In old coordinates it was written as
$\varepsilon=2(\cos k_xa+\cos k_ya+\cos k_za)$, and in new coordinates
$(x_1,x_2,x_3)$ it is 
$$
\varepsilon=2\cos k_2a+4\cos\frac{k_3a}{\sqrt{2}}\cos\frac{k_1a}{\sqrt{2}}.
$$

The equation (20) is a generalization of Harper's equation for the case
of magnetic field parallel to a diagonal of a square which is a face of the
elementary cell cube. Keeping a three-diagonal structure (20) has some
remarkable features with respect to standart Harper's equation.
First, (20) describes total (not "transversal"
$\varepsilon_{\perp}(k_1,k_3)$) energy as a function of quasimomentum
${\bf k}$: $\varepsilon=\varepsilon(k_1,k_2,k_3)$. Now $k_3=const$
corresponds to the plane perpendicular to ${\bf H}$ and $k_1=const$
describes the plane parallel to magnetic field. 
Second, (20) corresponds to anisotropic Harper's equation with anisotropy
ratio 
$$
4\cos\Big(\frac{k_3a}{\sqrt{2}}\Big)=\lambda\eqno(21)
$$
depending on the cross-section transversal to the magnetic field which is
fixed by choosing the value of $k_3$. It is well-known that Harper's equation
in the case $\lambda\ne 2$ describes 2D square lattice with anisotropic
transfer integrals [5,6]. According to (21) the persistent interval
for $\lambda$ is $-4\le\lambda\le 4$. In the quasiclassical limit it
describes the situation where open-type trajectories are available.
The contribution of open trajectories to the whole topological structure
varies with the respect to a cross-section $k_3=const$ being chosen. For
example, the cross-section $k_3=\pi\sqrt{2}/3a$ corresponds to $\lambda=2$
which is the case of isotropic Harper's equation and only close-type
orbits are avaliable. This is illustrated in Fig.3 where Fermi surfaces
(Fig.3a,b) and spectrum (Fig.3c) are shown for $p/q=31/60$. Then, the
cross-section $k_3=\pi/\sqrt{2}a$ gives us $\lambda=0$ which is a "full
anisotropic" limit for Harper's equation. Here as one can see in Fig.4a,b
only opened-type orbits are avaliable. It gives a continious energy spectrum
(Fig.4c). For remaining values of $\lambda$ both opened
and closed orbits are present. In the last case the spectrum
$\varepsilon_{\perp}$ consists of a continious and quasi discrete (narrow
subbands) parts.

As a conclusion, we suggested the expression for electron wave function which
satisfies to Peierls conditions and obtained Harper's equation from initial
principles of the tight-binding model. The Fermi surfaces are built
for different orientations of the magnetic field. In the following paper
we plan to study the electron states in 3D crystals in the case of the
magnetic field orientation parallel to the arbitrary translation vector
of the lattice. The properties of de Haas - van Alfen effect in this problem
will be studied.

One of us (V.Ya.Demikhovskii) wants to thank V.D.Selemir for constant support
and Prof. Michael von Ortenberg for numerous discussions of this problem.
This research was made possible due to financial support from the
Russian Foundation for Basic Research (Grant No. 98-02-16412).

\vspace{0.5cm}
{\bf Refernces}

\begin{enumerate}
\item M.Ya.Azbel', Zh. Exper. i Teor. Fiz. {\bf 46}, 929 (1964) [{\it in
Russian}].
\item D.R.Hofstadter, Phys.Rev.B {\bf 14}, 2239 (1976).
\item F.H.Claro, G.H.Wannier, Phys.Rev.B {\bf 19}, 6068 (1979).
\item P.G.Harper, Proc.Phys.Soc.(London) {\bf A 68}, 874 (1955).
\item H.Hiramoto, M.Kohmoto, Int.Journ.of Modern Physics {\bf 6}, Nos. 3\&4,
281 (1992).
\item M.Kohmoto, Y.Hatsugai, Phys.Rev.B {\bf 41}, 9527 (1990).
\item A.Barelli, R.Fleckinger, Phys.Rev.B {\bf 46}, 11559 (1992).
\item H.Silberbauer, J.Phys.: Condens. Matter {\bf 4}, 7355 (1992).
\item V.Ya.Demikhovskii, A.A.Perov, JETP {\bf 87}, 973 (1998)
\item V.Ya.Demikhovskii, A.A.Perov, Phys. Low-Dim. Structures {\bf 7/8}, 135
(1998).
\item D.Weiss, M.L.Roukes, A.Menschig, et al., Phys.Rev.Lett. {\bf 66}, 27
(1991).
\item T.Schl\"osser, K.Ensslin, J.P.Kotthaus, et al., Semicond. Sci. Technol.
{\bf 11}, 1582 (1996).
\item B.A.Boyko, A.I.Bykov, M.I.Dolotenko et al, book of abstracts, VIIIth
Int. Conference on Megagauss Magnetic Field Generation and Related Topics,
p.149, Tallahassee, USA.
\item R.E.Peierls, Z.Phys. {\bf 80}, 763 (1933).
\item V.Ya.Demikhovskii, A.A.Perov, D.V.Khomitsky, Proc. of the
VIIIth International Conference on Megagauss Magnetic Field Generation
and Related Topics (MEGAGAUSS-VIII), Tallahassee, USA (1998) (in press).
\item J.Zak, Phys.Rev.A {\bf 134}, 1602 (1964);~Phys.Rev.A {\bf 134}, 1607 (1964).
\end{enumerate}
\end{document}